%% file: main.tex
\documentclass[reprint,
superscriptaddress,amsmath,amssymb,aps,prl,floatfix]{revtex4-2}
\usepackage[utf8]{inputenc}
\usepackage{xspace}
\usepackage{dcolumn}
\usepackage{textcomp}
\usepackage{bm}
\usepackage{geometry}
\usepackage{float,subcaption}
\usepackage{tikz}
\usepackage{booktabs}
\usetikzlibrary{positioning}
\usepackage{amsmath,amssymb}
\usepackage{graphicx}
\usepackage{xcolor}
\DeclareUnicodeCharacter{0301}{\'{e}}

\begin{document}
\input{title}

\begin{abstract}
{Ensuring solid-state lithium batteries perform well across a wide temperature range is crucial for their practical use. Molecular dynamics (MD) 
simulations can provide valuable insights into the temperature dependence of the battery materials, however, the high computational cost of 
\textit{ab initio} MD poses challenges for simulating ion migration dynamics at low temperatures. To address this issue, accurate machine-learning interatomic potentials were trained, which enable efficient and reliable simulations of the ionic diffusion processes in Li$_6$PS$_5$Cl over a large temperature range for long-time evolution. Our study revealed the significant impact of subtle lattice parameter variations on $Li^+$ diffusion at low temperatures and identified the increasing influence of surface contributions as the temperature decreases. Our findings elucidate the factors influencing low-temperature performance and present strategic guidance towards improving the performance of solid-state lithium batteries under these conditions.}
\end{abstract}
\maketitle

\textit{Introduction}.-A long-standing challenge for developing lithium ion batteries (LIBs) is to ensure their stable performance over a wide temperature range. However,
in the low-temperature regime, the LIBs encounter a series of challenges, including a decline in ionic conductivity, dendrite growth, and electrolyte freezing. In the past, various strategies have been tried to enable the use of conventional lithium-ion batteries with liquid organic electrolytes at low temperatures. These include i) incorporating additives into the electrolyte to decrease the $Li^+$ desolvation energy and enhance the transference number of lithium ions \cite{zhao2022insight}; ii) producing low-resistance SEI components \cite{kim2023interface,song2021new} and altering the microstructure of electrode materials \cite{lee2021outstanding} to reduce the migration barrier of lithium ions; and (iii) selecting electrolyte solvents with appropriate melting points to overcome freezing issues of the electrolyte at low temperatures \cite{fathollahi2019use}, to name a few.

The solid-state lithium batteries are regarded as a promising candidate for the next-generation rechargeable techniques because of their potentials to achieve high energy density and safety \cite{xu2018interfaces,kamaya2011lithium,han2016electrochemical,auvergniot2017interface,zhang2018new,he2019crystal}. However, the substitution of liquid electrolytes with solid-state ones poses further challenges in improving the 
low-temperature performance because of the poor understanding of the complex migration mechanism in both bulk and surface/interface area within different temperature ranges. The challenges exist in both experimental measurements and theoretical simulations. In the experimental case, the dynamic changes during cycling procedures impose high demands on the real-time and in-situ capabilities of laboratory approaches, while for theoretical simulations the challenge
arises from the strong rare-event nature of the ion migration phenomenon at low temperatures and the limitations from the expensive \textit{ab initio} calculations \cite{he2018statistical}. In essence, to simulate battery materials at low temperatures, a theoretical approach which can handle large-sized systems containing both bulk and surface/interface and allow for long-time evolution with an accuracy close to \textit{ab initio} molecular dynamics (AIMD) \cite{car1985unified} is indispensable.

So far, there have been very few attempts in this regard. Heo et al. simulated the effect of microstructures on ion transport using the phase-field approach in the large system composed by Li$_{7-x}$La$_3$Zr$_2$O$_{12}$ (LLZO) garnet solid electrolyte particles \cite{heo2021microstructural}. Baktash et al. demonstrated that \textit{ab initio} nonequilibrium molecular dynamics enables reliable estimates of the diffusion coefficients even with limited simulation duration \cite{baktash2020diffusion}. In addition to the above attempts, via the application of deep neural networks, machine-learning interatomic potentials (MLIPs) have exhibited promising capabilities in tackling this issue \cite{miwa2017interatomic,miwa2018molecular,li2017study,huang2021deep,behler2016perspective,noe2020machine,behler2007generalized}. In this work, we explore the vast possibility enabled by the MLIPs by looking at the Li ion transport in one of the prototypical solid-state electrolyte materials, argyrodite Li$_6$PS$_5$Cl (LPSC). This effort leads to novel mechanistic insights into the ion transport in this material, 
especially at low temperature range.

In solid-state electrolytes, the sulfides always show higher ionic conductivity than oxides because of the weaker binding of Li-S bonds. Among them, LPSC has been extensively investigated due to its high potential technological importance, arising from its high ionic conductivity and low temperature processability \cite{deiseroth2008li6ps5x}. Interestingly, researchers have revealed that there exists three types of ion migration events in the system, named as ``doublet jump", ``intracage  hoppings", and ``intercage hoppings". These three processes exhibit varying energy barriers and are activated at different temperatures \cite{de2016diffusion}, rendering it an excellent prototype material for investigating the impact of temperature on the performance of solid-state lithium batteries, with insights gained at the atomic scale, particularly within the low-temperature range.

In this work, we start by training a machine-learning model to obtain accurate interatomic potentials for the Li-P-S-Cl system, containing both
bulk and surface structures in the training dataset, which enabled
us to run MD simulations for large models up to 832 atoms and long duration up to 10 nanoseconds (ns) at various temperatures. By analyzing the
MD data, we observed distinctive ion migration behaviors under high and low-temperature conditions. Our findings reveal that even minor alterations in lattice parameters exert a substantial influence on ionic conductivity at low temperatures, while their impact is negligible at elevated temperatures. Through simulations of ion transport in a model incorporating surfaces, we identified the predominance of surface transport at lower temperatures in LPSC.

\textit{Theoretical methods and computational details}.-The primary requirement for training a high-quality MLIP is to have a large and accurate dataset from first-principles calculations. To this end, the Atomic-orbital Based Ab-initio Computation at UStc (ABACUS) \cite{chen2010systematically,li2016large} software package is used to create energies and forces encompassing diverse chemical environments in LPSC. Specifically,  the Perdew-Burke-Ernzerhof (PBE) \cite{perdew1996generalized} generalized gradient approximation is adopted for the exchange-correlation functional of DFT, and the linear combination of numerical atomic orbital (NAO) basis sets at the level of double-zeta plus polarization (DZP) as developed in Ref.~\cite{Lin2021PRB} 
are used in all PBE calculations.  Furthermore,
the multi-projector ``SG15-ONCV"-type norm-conserving pseudopotentials are employed to describe the interactions between nuclear ions and valence electrons. 
Previously, ABACUS with such settings has been successfully used to study aluminum-ion battery systems \cite{ABACUS_graphite,Wang/etal:CPB:2021}. 
In the present work,  the electronic convergence is gauged against a criterion of $10^{-8}$ eV; a $3\times 3 \times 3$ {\bf k}-mesh is used for
the conventional unit cell, and a $\Gamma$-only {\bf k}-point is used for the $2\times 2\times2$ supercell, respectively. 
Stretch and compression from the equilibrium lattice up to $5\%$ have been introduced to model the atomic environment under stress. The simulation of DFT is divided into two parts. The first part is to use NVT ensembles to perform 10ps and 1ps AIMD calculations on the unit cell and $2\times 2\times2$ supercell respectively as the initial training set. Simulations are conducted for temperatures ranging from 800 K to 1200 K with an interval of 200 K. The second part is the static calculations of the selected structures in each DP-GEN iteration(iteration details are provided in the Supplemental Material(SI) Table.~S1)), which are created by the MD simulations of NVT and NpT ensembles using MLIP from 200 K to 1200 K.

The Deep Potential Generator (DP-GEN) \cite{zhang2019active,zhang2020dp} is employed to acquire a minimal set of training data via an efficient and sufficient sampling process, thereby guaranteeing the attainment of a reliable potential energy surface. The DeePMD-kit \cite{zhang2018deep,kingma2014adam,wang2018deepmd} is employed to generate a uniformly accurate MLIP, which is then utilized for simulating larger systems with longer duration in dynamic evolution processes through the LAMMPS software \cite{plimpton1995fast}.  To train the MLIP with DeePMD-kit, the two-atom embedding descriptor \text{``se\_e2\_a"} is adopted to construct the information of atomic configurations, including all information (both angular and radial) of atomic configurations. The number of neurons in each hidden layers of the embedding net are ``240,240,240". the cutoff radius $r_c$ is set to $8.0$ \AA, and the $r_{sc}$ value of $0.8$ \AA\ is used to determine where to start smoothing \cite{zhang2018end}. Such dimension setting are sufficiently expansive to encompass Li ions within the second coordination shell in Li$_6$PS$_5$Cl. The lower and upper bound of the selection criteria of the deviation of forces ($\sigma_{f}$) as 0.2 eV/\AA{} to 0.4 eV/\AA. Each iteration is trained for 400000 steps, following 21 iterative explorations under the NpT and NVT ensembles from 200 K to 1200 K.  The surface information is included in the last 7 iterations by adding DFT simulation results for Li$_6$PS$_5$Cl(001) surface into the training data. A converged model is thus attained, with root mean square error (RMSE) less than 2.08 meV and 0.06 eV/\AA{} for energy and force, respectively. Further details are provided in the SI (Fig.~S1, S2 and Table.~S1). The obtained MLIP including surface information is employed in the following kinetic simulations by LAMMPS code. The diffusion coefficients at each temperature are determined through Mean Squared Displacement (MSD) data of Li$^+$ ions by applying the Einstein relation \cite{he2018statistical}.

\textit{The Effectiveness of DPMD simulations}.-The essential aspect of the present research is to make sure that the trained MLIP is sufficiently accurate, which can be judged by several criteria.  The first necessary criterion is the small enough RMSE of the trained model 
in the energies and forces, as
indicated above. In addition, the nearly perfect agreement between the equation of state (EOS) curves as yielded by DFT and MLIP 
also indicates the high accuracy of the trained deep-potential (DP) model (Fig.~S1(b)). Furthermore, to check if the MLIP accurately captures the ionic transport dynamics of the system as well, we calculate the diffusion coefficients and fit the Arrhenius plots, from which the Li$^+$ ionic conductivity and activation energy from both AIMD and deep-potential MD (DPMD) simulations can be extracted. The MSD results of two methods 
obtained within the range from 700 K to 1200 K are illustrated in Fig.~S3. Due to the high computational cost of AIMD, the diffusion process simulation is conducted for 20 ps, in which the initial 5 ps are regarded as the time required for the system to reach equilibrium at the simulation temperature, and the subsequent 6-20 ps are used for statistical analysis of the MSD curves. To avoid different statistical errors due to varying data volumes, we employ the same model structure, simulation duration, and MSD statistical method for both AIMD and DPMD simulations in the comparative analysis. In Fig.~S3(c), the close agreement between the blue and orange inverted triangles indicates the consistency of diffusion coefficients extracted from AIMD and DPMD simulations for the unit cell. This confirms that the accuracy of DPMD calculations is close to the AIMD and the former is suitable for the following investigations.

\begin{figure}[hpt]
\centering
\includegraphics[width=1.0\columnwidth]{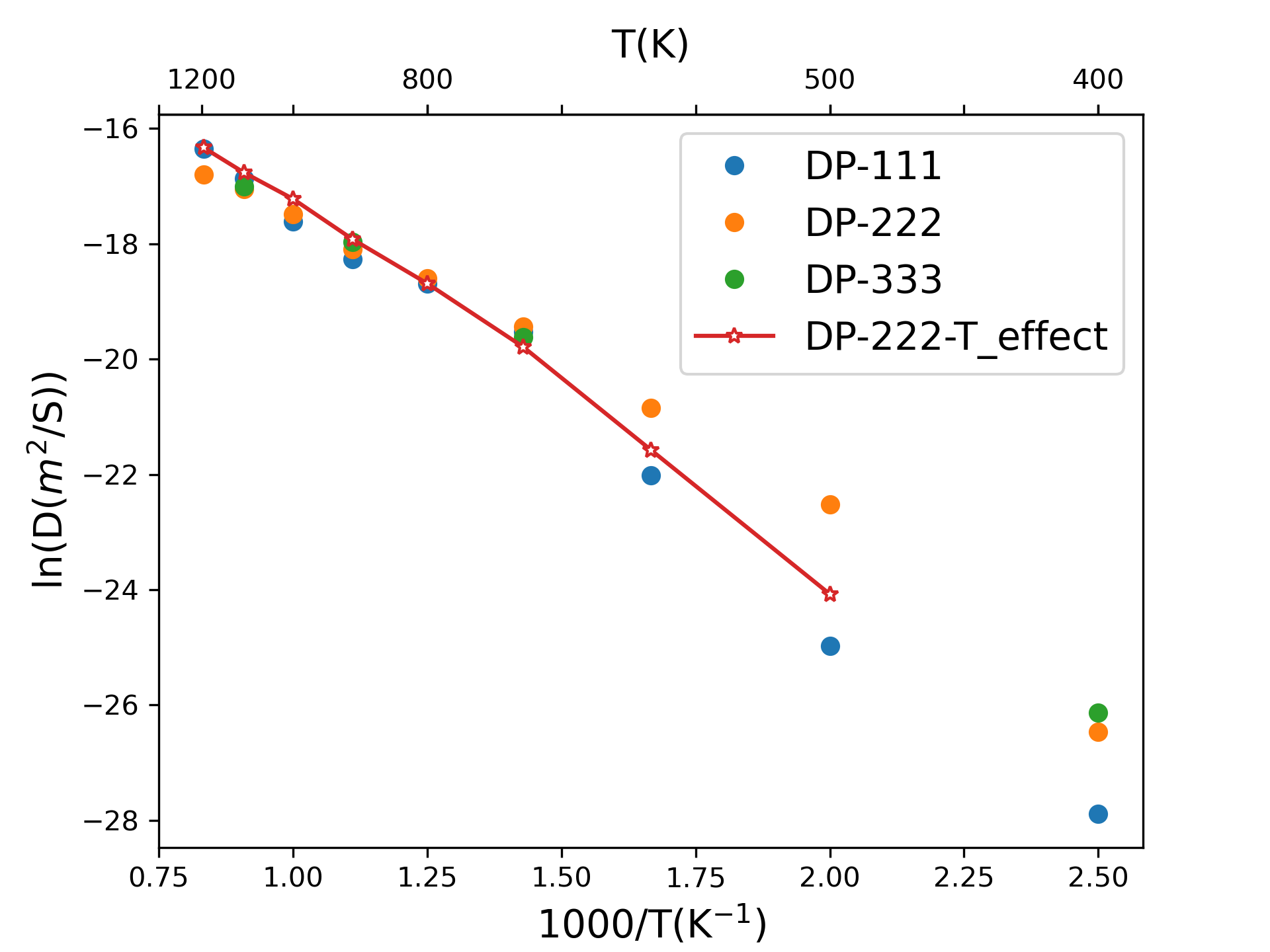}
\caption{\label{fig:lnD-1/T}Arrhenius plots of Li$^+$ diffusion coefficient $D$ as a function of temperature $T$ in LPSC obtained from AIMD and DPMD simulations. The blue, orange, and green dots represent DPMD results from simulations for the unit cell, $2\times 2\times2$ and $3\times 3\times3$ supercells, respectively.  The red stars represent the $2\times 2\times2$ supercell results,  taking the effect of thermal expansion into account. The simulations for $3\times 3\times3$ supercell performed only on 4 temperature points because of the increasing computational cost.}
\end{figure}

By utilizing the MLIP, the MD simulations can be easily extended to the time scale of nanoseconds. Longer simulation time enables the accumulation of more ionic migration events, providing sufficient data for statistically analyzing the MSD even at low temperatures. The blue points in Fig.~\ref{fig:lnD-1/T} represents the result of DPMD of 1 ns. At temperature as low as 400 K, the simulation time is extended to 10 ns to obtain sufficient statistical events. The simulation at 300 K for 100 ns still shows no hopping events, indicating that the simulation over much longer time scale is required to study the room-temperature kinetics.

The DPMD method is also employed for larger models to eliminate the finite-size effect in the simulations. The orange and green dots in Fig.~\ref{fig:lnD-1/T} show the diffusion coefficients extracted from DPMD simulations for $2\times 2\times2$ supercell and $3\times 3\times3$ supercells, respectively. The detailed MSD data are illustrated in Fig.~S4. The close agreement between the two sets of data indicates that the size effects in LPSC are not very pronounced, and the results are well converged with 
$2\times 2 \times 2$ supercell. This is possibly due to the fact that each intracage migration primarily occurs in a localized region of 1/8 unit cell, while the intercage migrations take place over relatively short distances between neighboring cages. Both of them are highly localized events, thus allowing for accurate statistical analysis even without invoking very large supercells. To balance the accuracy and computational cost, the subsequent discussion on the kinetic properties are mainly based on the simulations of $2\times 2\times2$ supercell.

 \textit{Temperature Effects on Transport Behaviors}.- As the temperature increases from low to high, the thermal vibrations of lattice atoms intensify, leading to a series of possible changes including the variations of lattice parameters, thermal expansion coefficients, degrees of lattice disorder, phase transitions, etc. Almost all of these factors are closely associated with the ion migration phenomenon. We carried out DPMD simulations with the NpT ensemble in LPSC unit cell for 1 ns and average the cell volume over the last 0.5~ns to obtain the lattice parameters and thermal expansion coefficients for each simulated temperature. Detailed results are shown in Fig.~S5. As Li$_6$PS$_5$Cl crystallized in the cubic F$\overline{4}$3m space group, we applied the ``$a$=$b$=$c$"  
 constraint for the lattice constants during the simulations. In the temperature ranging from 100 K to 1200 K, the lattice constants increase from 10.05~\AA\ to 10.56~\AA, with an increment of less than 3$\%$. 

 To investigate the temperature effects on ionic conductivity, DPMD simulations with NVT ensemble are conducted for $2\times 2\times2$ supercell at various temperatures. The chosen cell parameters depend on the temperature in a way as given in Fig.~S5. The extracted diffusion coefficients are shown in Fig.~\ref{fig:lnD-1/T} as red stars. The lattice parameter adopted in the simulations without considering the thermal expansion is 10.279~\AA, and the corresponding results are represented by orange dots in Fig.~1. The temperature effects cause the lattice expanding to 10.369~\AA\ at 1000~K and contracts to 10.130~\AA{} at 400~K. Sufficient number of statistical hopping events are obtained for temperatures above 500~K, however, no Li $^+$ ion migration event is recorded even for simulation time of 10~ns at 400~K, the statistic is not good enough for 400 K, and hence the diffusion coefficient cannot be obtained. When the temperature is above 800~K, despite the gradual increase in lattice parameters, the obtained 
 Li $^+$ ion diffusion coefficients show little dependence on the lattice parameters, as indicated by the close agreement between orange dots and red stars
 in Fig.~\ref{fig:lnD-1/T}. In contrast, the reduction of lattice parameters in lower temperature regime leads to a pronounced decrease in the ion diffusion coefficients. This indicates that the usual procedure, i.e., fitting directly the data within the high-temperature range to extrapolate the ionic conductivity at the room temperature, can introduce non-negligible errors due to neglecting the temperature dependence of the lattice parameters. For example, by fitting five data points from 800~K to 1200~K and extrapolating to room temperature, we obtain an ionic conductivity of 2.20$\times 10^{-3}$ S/cm, while by fitting data from 500~K to 700~K, the ionic conductivity is estimated to be 2.16$\times 10^{-4}$ S/cm at 300 K. In previous reports \cite{boulineau2012mechanochemical,baktash2020diffusion,hanghofer2019fast,rayavarapu2012variation,deng2017data,yubuchi2015preparation,yu2016synthesis}, the conductivity of LPSC obtained by various experiments with the values around 1.1$\times 10^{-3}$ to 6.0$\times 10^{-5}$ S/cm. It is worth noting that both the phase transition \cite{kong2010lithium} and the change of transport mechanisms \cite{de2016diffusion} are possible origins for the slope turning point exhibited in the diffusion coefficient curves. However, the slope variation observed here in LPSC is merely caused by the slight decrease of the lattice parameter because of the temperature effects.

  In addition to the necessity of conducting MD simulations in relatively low temperature range to obtain reliable ionic conductivity at temperatures lower than 300 K, the sensitivity of ionic migration to cell volume also reflects that the lattice parameters play a more crucial role at low temperatures than at high temperatures. As the temperature decreases from 700~K to 400~K, the discrepancy between the diffusion coefficients extracted for considering with and without thermal expansion in $2\times 2\times2$ supercell is increasing gradually. At 400~K, the simulations with subtle smaller lattice parameters even found no hopping events.
  
\begin{figure}[ht]
    \centering
    \includegraphics[width=1.0\linewidth]{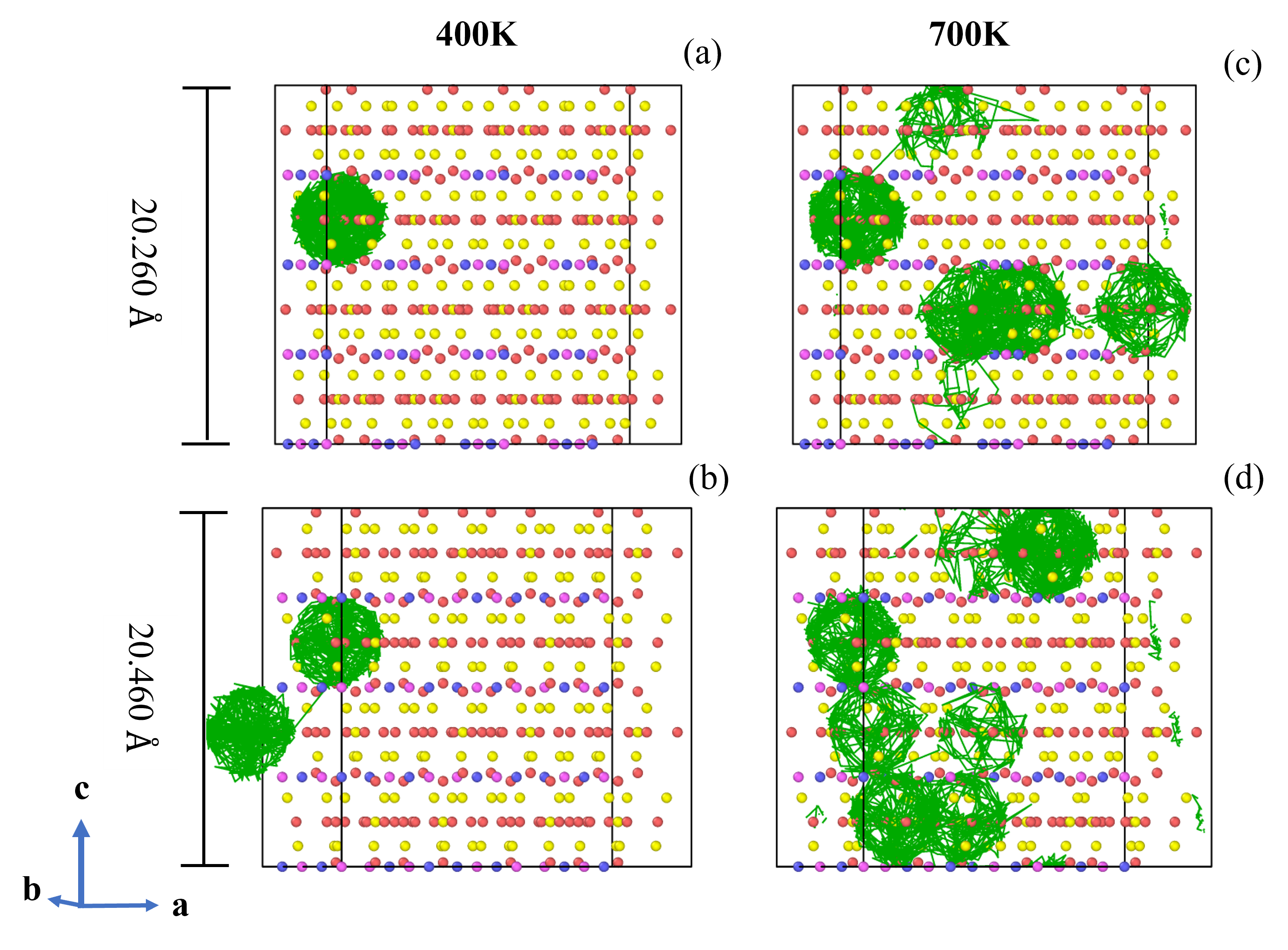}
    \caption{ The motion trajectory (green lines) of the Li ion with maximal MSD in 3~ns at 700~K for lattice parameter of (a) 10.130 \AA\ and (b) 10.230 \AA, and in 10~ns at 400~K for lattice parameter of (c) 10.130 \AA\ and (d) 10.230 \AA. The trajectories are plotted for $2\times 2 \times 2$ supercells. The colors of Li, P, S, and Cl are red, blue, yellow, and pink, respectively.}
    \label{fig:traj-4_700K}
\end{figure}

  DPMD simulations for two models with lattice constants of 10.130 \AA\ and 10.230 \AA are performed at both 400~K and 700~K for comparison. The variation of the ion migration ability is also manifested in the trajectories of Li ions as shown in Fig.~\ref{fig:traj-4_700K}. The green lines are the connections between two positions with time interval of 10 ps, and the trajectories of the Li ion with maximal MSD in each case are illustrated. The connection among different cages during the simulations occurs within 3~ns for both of the two models at 700~K, while the migration between neighboring cages only happens in model with larger lattice parameter at 400~K even with longer simulation time of 10 ns (as shown in Fig.~S6). The lattice stress of the two models is estimated via the DFT calculations for the two initial structures. The results indicate that, as the lattice parameter changes from 10.130 \AA\ to 10.230 \AA, the total stress decreases from 23.54 kBar to 9.60 kBar. It suggests that stress regulation at low temperatures can be an efficient way to adjust the ionic conductivity. This scenario can be elucidated by the different triggering conditions for intracage and intercage diffusions. The intracage migration happens at the smooth potential energy surface with very low energy barrier around the 4c-site $S$ atoms, which can be activated even at low temperature. However, the intercage migration with higher energy barrier is the rate-determining step for Li$^+$ ion diffusion in LPSC and only happens in the environment with enough space or higher temperature. The former increases the driving force by providing more thermal vibration energy and the latter decreases the energy barrier by providing wider diffusion channel. At high temperatures, both two migration behaviors are active, and thus there is negligible effect by lattice parameters. While at low temperatures, the subtle enlargement of the lattice vary the barrier height hence stimulating more intercage migration modes. 

 We conduct an extensive statistical analysis focusing on the frequency of cage-to-cage transport events for Li$^+$ at various temperatures, as well as the count of Li$^+$ involved in these occurrences. The intercage migration events can be identified by jumps between plateaus observed in the MSD as shown in Fig.~S7. The number of intercage Li$^+$ hopping and the time of the first hopping event are summarized in Table \ref{tab:widgets}. For temperatures below 600~K, there is a relatively modest occurrence of transfers and a limited count of Li$^+$ involved in cage-to-cage transport. In this temperature range, the lattice parameter change introduced by the low temperature effect will influence the ion migration significantly. However, upon reaching temperatures of 700~K and 800~K, the transport of Li$^+$ between cages experiences a significant increase. The hops between cages are already activated for all the Li$^+$ ions. As a result, the lattice influence the kinetic properties weakly.
\begin{table*}
\caption{\label{tab:widgets}
 		{The time interval required for Li$^+$ ions to initiate intercage migration (denoted by $t_{init}$), the count of atoms involved in intercage migration ($N_{atom}$), the average number of intercage transfers per atom ($N_{transfers}$), and the average residence time ($t_{residence}$) within each cage within an 8 ns timeframe across temperatures (T) spanning from 400~K to 800~K, for the simulations of the $2 \times 2 \times 2$ supercell.}}
\resizebox{\linewidth}{!}{
\begin{tabular}{p{2cm}p{2cm}p{2cm}p{2cm}p{2cm}}
\toprule
\textbf{$T (K)$} & \textbf{$t_{init} (ps)$} & \textbf{$N_{atom}$} & \textbf{$N_{transfers}$} & \textbf{$t_{residence} (ns)$} \\ 
\midrule
400 &  4091 & 2 & 1  & \textgreater 8 \\ 
450 &  1292 & 22 & 1  & \textgreater 8  \\
500 & 453 & 85 & 2  & 7   \\
550 & 205 & 173 & 2  & 5  \\
600 &  66.6 & 192 & 4  & 2\\ 
700 & 32 & 192 & 12  & 0.7 \\ 
800 & 4 & 192 & - & - \\ 
\bottomrule
\end{tabular}
}
\end{table*}

\begin{figure*}
    \begin{subfigure}{0.49\linewidth}
        \centering
        \includegraphics[height=5cm]{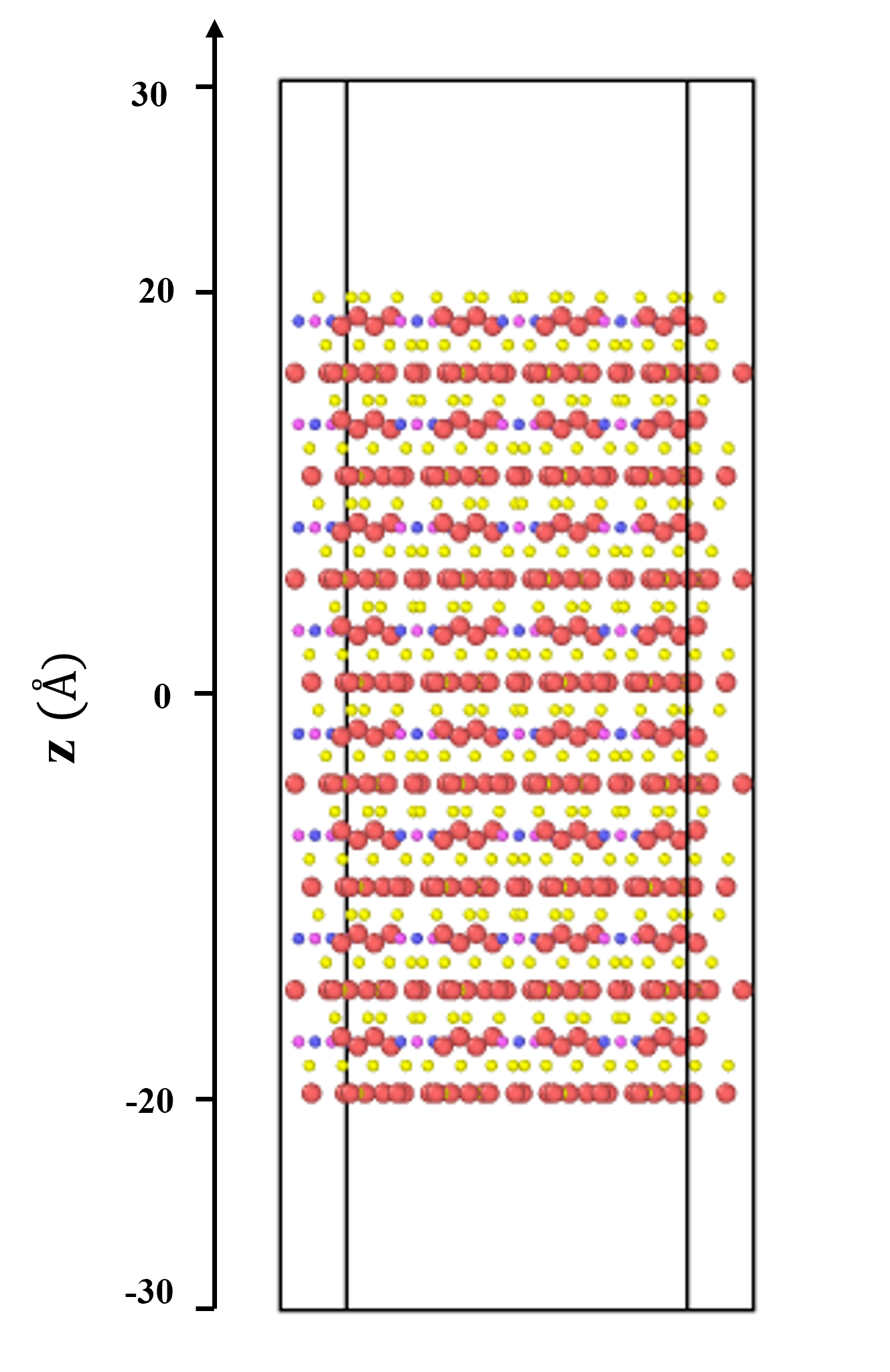}
        \caption{}
        \label{fig:GB-struYZ}  
    \end{subfigure} 
        \begin{subfigure}{0.49\linewidth}
        \centering
        \includegraphics[height=5cm]{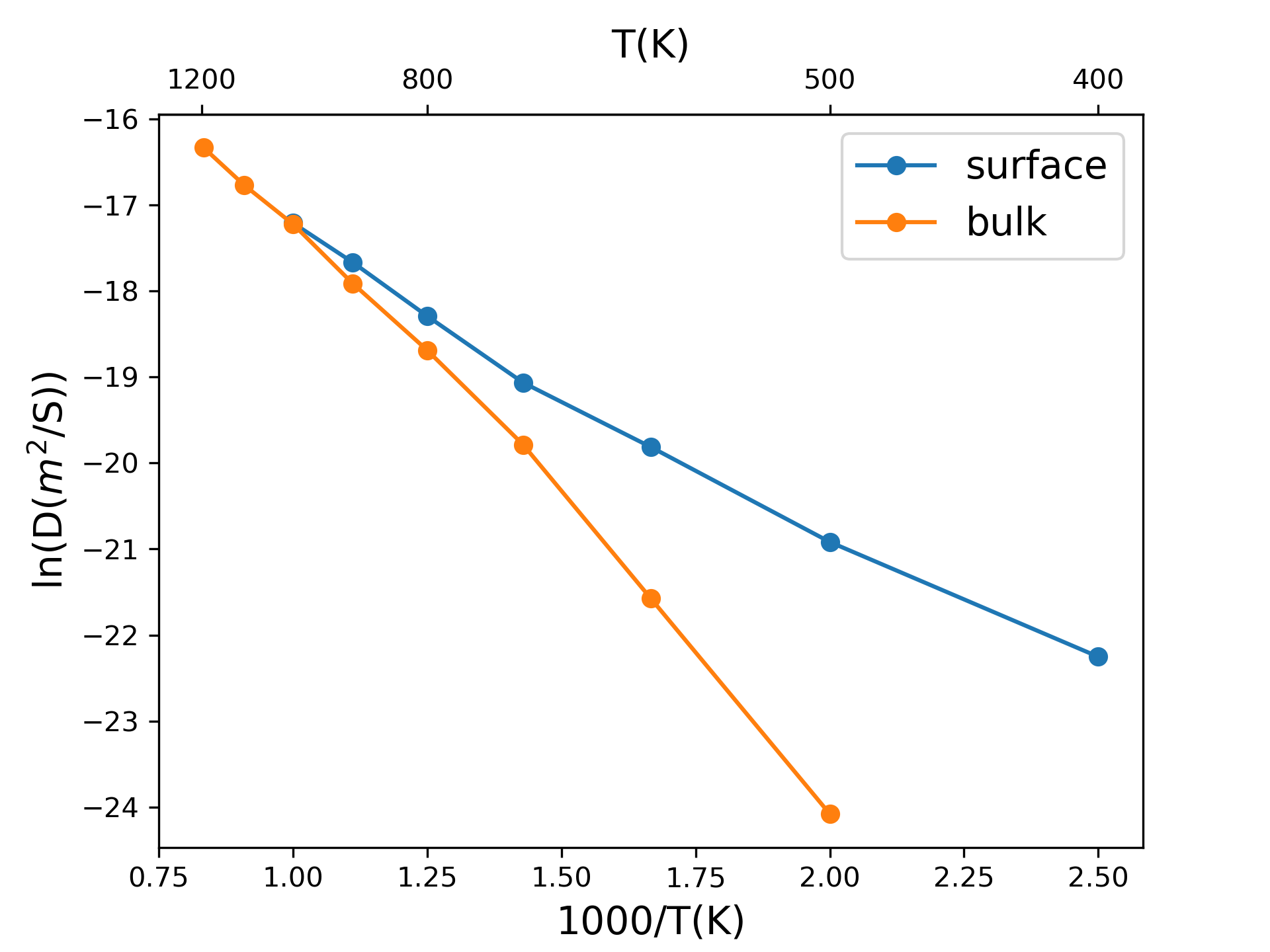}
        \caption{}
        \label{fig:lnD-sur-bulk2}  
    \end{subfigure} 
        \begin{subfigure}{1.0\linewidth}
        \centering
        \includegraphics[width=1.0\linewidth]{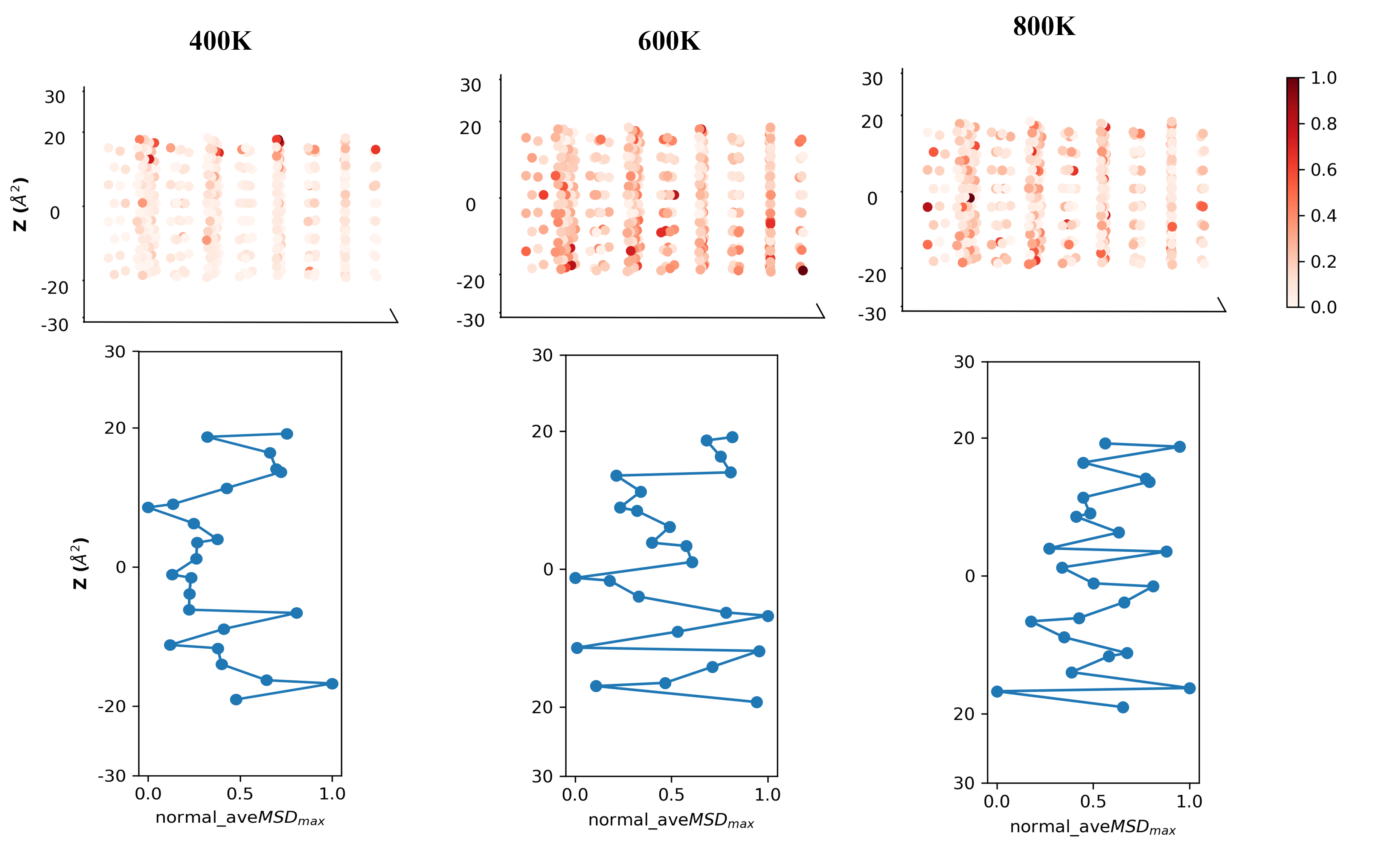}
        \caption{}
        \label{fig:max-msd-3d}  
    \end{subfigure} 
    \caption{(a) Side view of the surface model, with a vacuum thickness of 20~\AA. (b) Comparison of diffusion coefficients in bulk and surface models at different temperatures. (c) The maximum MSD value of each Li$^+$ ion represented by white to red at upper figures, and the lower figure visualizes the average maximum MSD value for each row, all the MSD values normalized to the maximum one in each case.}
    \label{GB}
\end{figure*}

\textit{Surface Transport Characteristics}.-Material surfaces, due to their distinct chemical environments compared to the bulk, may exhibit different ionic transport characteristics\cite{golov2021molecular,shi2012direct,wang2018review}. To detect the transport at the surface region, we construct a slab model containing two (001) surfaces, as illustrated in Fig~\ref{fig:GB-struYZ}. The layers near the center of the 
slab mainly represent the bulk properties, whereas the layers close to z= $\pm 20$ \AA\ signify the surfaces. The thickness of the vacuum between two surfaces is 20~\AA. The model contains 832 atoms in total, of which 384 are Li$^+$ ions. The top layer is the PS$_4$-terminated surface, while the bottom layer is the Li-terminated surface. The DPMD simulations with 1-10 ns are carried out for this system. The lower the simulation temperature, the longer the simulation duration is adopted. The comparison of the Arrhenius relationship between the bulk and surface models is given in Fig~\ref{fig:lnD-sur-bulk2}. The two curves show significant differences at low temperatures, and as the temperature increases, the difference diminishes and the diffusion coefficients converge to the same values at 1000~K. It can be inferred that the Li$^+$ ions near surfaces are more mobile than the ions in the bulk region at 400 K, because the diffusion coefficient evaluated for the surface model is much larger than that of the bulk model. The maximum MSD values for each Li$^+$ ion are calculated from the trajectories of DPMD and displayed by the colormap in Fig~\ref{fig:max-msd-3d}. At 400~K, the ions with large MSD represented by red dots are mainly distributed at the upper surface. At 800~K, the red dots appear also in bulk region, indicating the contributions to ionic diffusivity from surface and bulk regions become similar. Thus the importance of surface ionic migration gradually increases as temperature decreases from 800~K, through 600~K to 400~K.

 The mechanism of the above phenomenon can be attributed to the loose bonding environment of the surface part. Because of the existence of vacuum space, the atoms near 
 the surfaces can move towards vacuum space to expand the lattice. At 400~K, the distances between the 1st and 2nd S layer near the surfaces range from 7.4 to 8.5~\AA\, which is larger than the S-layer distance of 7.2~\AA\ in the bulk region. The reason for the increase of ion migration at surfaces is consistent with the influence of
lattice parameter discussed in section of thermal effect for the bulk system, in which a subtle increase of the lattice parameter is beneficial to the Li$^+$ ion migration, too. It provides possible strategies to improve the low temperature ion migration by adjusting the surface area, e.g. by doping or coating for the surfaces, introducing the second phase particles, or increasing the surface ratio. However, the bond breaking is not sufficiently considered in current surface model because it is simplified by cutting between the Li layer and the PS$_4$ layer without breaking any P-S bond. The bonding and merging of different surfaces to form interface are still complex processes that need to investigate in the future work.

 \textit{Conclusions}.-By constructing a MLIP for Li-P-S-Cl system, we realize MD simulations of solid-state electrolytes with long-term evolution over 
 a wide temperature range. The Sufficient Li$^+$ ion migration events are obtained for statistical analysis for temperatures as low as 400~K, thanks to  the acceleration of the simulations with MLIP. It provides the chances to discern the temperature effect on ionic transport behavior in Li$_6$PS$_5$Cl. By exploring the influence of lattice expansion on diffusion coefficient, accounting for simulation time, finite-size effects, and thermal expansion factors, confirming the effectiveness of DPMD simulations and ensuring the necessity of estimated ionic conductivity using low temperature simulations with enough hopping events.

 A turning point is found around 600~K below which the intercage hopping is not fully activated even for simulations lasting for 10~ns, resulting that the effect of subtle change of lattice parameters in the low-temperature regime on ion migration is more significant than at high temperatures. Besides, the modeling for system with surfaces indicates that as the temperature decreases, the contribution from the surface Li$^+$ ion migration is enhanced and the underlying reason is attributed to the loose bonding environment near the surfaces. The simulation method in this work realizes the correlated research of the kinetic properties in SSEs at both high and low temperatures, which help to identify the influencing factors hard to find in AIMD simulations because of the limited simulation time-scale. The significant contribution of surface transport at low temperatures discovered in this study provide strategies for enhancing the low-temperature ionic conductivity of solid-state electrolytes. Changing lattice through bulk doping and through coating or introducing second phase particles are potentially effective approaches to improve ionic transport performance at low temperatures. It also makes possible to achieve better kinetic properties at low temperatures by increasing the proportion of surface area. It provides a feasible way to understand the low temperature performance of all solid-state lithium batteries by simulating more complex large-scale surface/interface models at low enough temperature with long time evolution.

 \section{Acknowledgment}
The authors would like to thank Xinzheng Li, Qijun Ye, Jun Cheng, and Han Wang for helpful discussions. 
This work was supported by the Strategic Priority Research Program of Chinese Academy of Sciences 
(Grant No. XDB0500201), and by the National Natural Science Foundation of China (Grants Nos. 52172258, 52022106, 
12134012, 12374067, and 12188101).  The numerical calculations 
  in this study were partly carried out on the ORISE Supercomputer. This computing resource was also provided by the Bohrium Cloud Platform (https://bohrium.dp.tech), which is supported by DP Technology.
\bibliographystyle{apsrev4-2.bst}
\bibliography{sample.bib}

\end{document}


\title{Supplementary Information of ``Mechanistic Insights into Temperature Effects for Ionic Conductivity in Li$_6$PS$_5$Cl''}
\author{Zicun Li}
\affiliation{College of Physics
Nanjing University of Aeronautics and Astronautics (NUAA)
Nanjing 211106, China}
\affiliation{Beijing National Laboratory for Condensed Matter Physics, Institute of Physics, Chinese Academy of Sciences, Beijing 100190, China}
\author{Jianxing Huang}
\affiliation{State Key Laboratory of Physical Chemistry of Solid Surfaces, iChEM, College of Chemistry and Chemical Engineering,
Xiamen University, Xiamen 361005, China}
\author{Xinguo Ren}\email{renxg@iphy.ac.cn}
\affiliation{Beijing National Laboratory for Condensed Matter Physics, Institute of Physics, Chinese Academy of Sciences, Beijing 100190, China}
\author{Jinbin Li}\email{jinbin@nuaa.edu.cn}
\affiliation{College of Physics
Nanjing University of Aeronautics and Astronautics (NUAA)
Nanjing 211106, China}
\author{Ruijuan Xiao}\email{rjxiao@iphy.ac.cn}
\affiliation{Beijing National Laboratory for Condensed Matter Physics, Institute of Physics, Chinese Academy of Sciences, Beijing 100190, China}
\author{Hong Li}
\affiliation{Beijing National Laboratory for Condensed Matter Physics, Institute of Physics, Chinese Academy of Sciences, Beijing 100190, China}

\maketitle
\onecolumngrid

\section{Accuracy test of DP models}
\begin{figure}[ht]
\centering
    \begin{subfigure}{0.6\linewidth}
        \centering
        \includegraphics[width=1.0\linewidth]{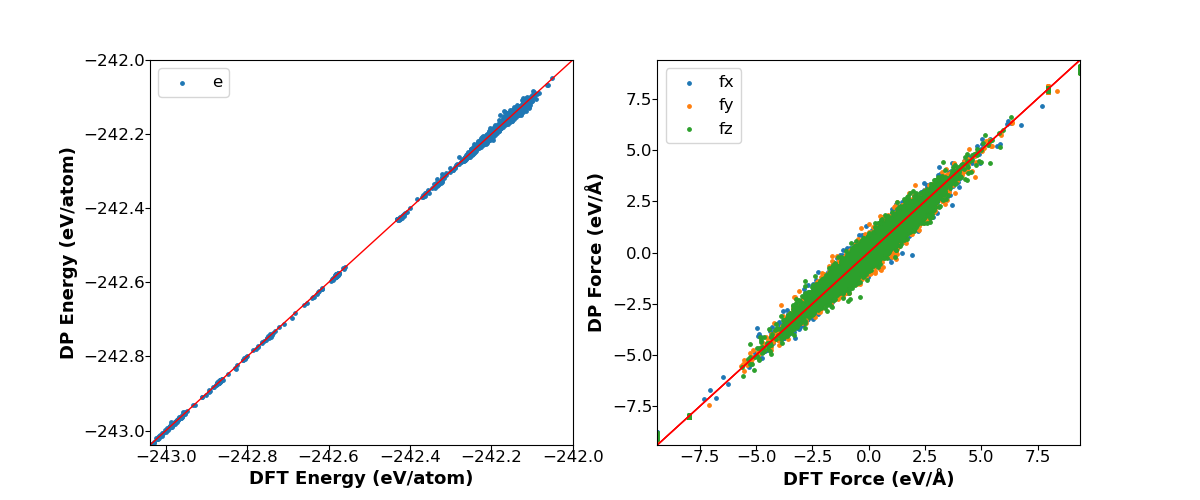}
        \caption{}
        \label{fig:RMSE}  
    \end{subfigure} 
    \begin{subfigure}{0.35\linewidth}
        \centering
        \includegraphics[width=1.0\linewidth]{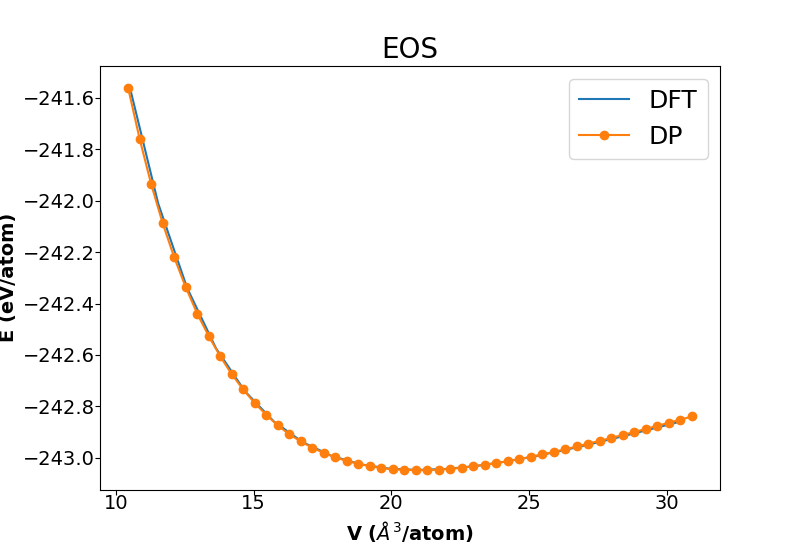}
        \caption{}
        \label{fig:EOS}       
    \end{subfigure}
  \caption{ (a)Comparison of the energy and force computed by DFT and DP model for Li$_6$PS$_5$Cl; (b) Comparison of the equation of states (EOS) curves of the DFT calculations (full lines) and the predictions by the DP model (dashed lines).}
  \label{fig:rmse-eos}
\end{figure}

\begin{figure}[htbp]
    \centering
    \begin{subfigure}{0.45\linewidth}
        \centering
        \includegraphics[width=1.0\linewidth]{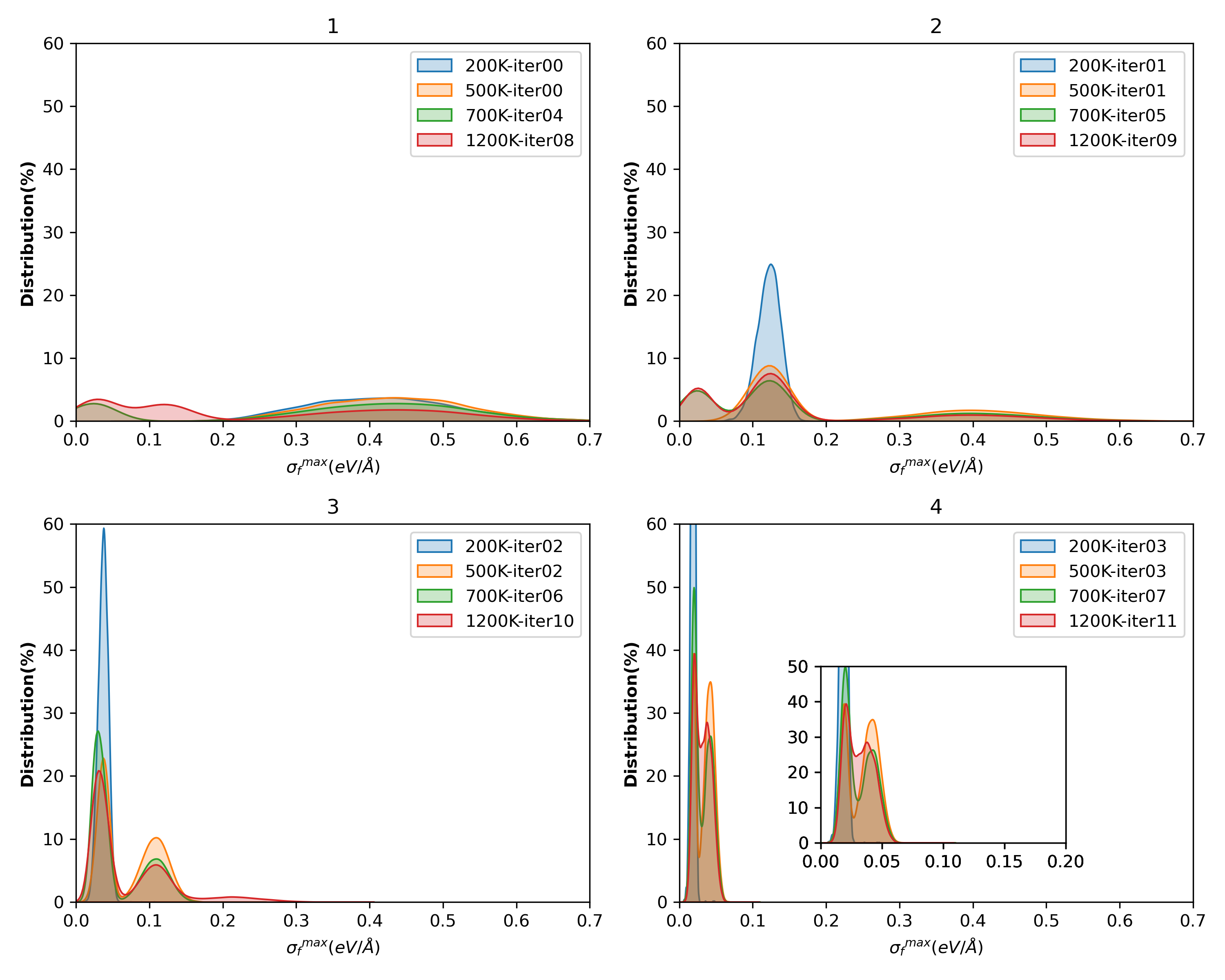}
        \caption{}
        \label{fig:max_devi}  
    \end{subfigure} 
    \begin{subfigure}{0.45\linewidth}
        \centering
        \includegraphics[width=1.0\linewidth]{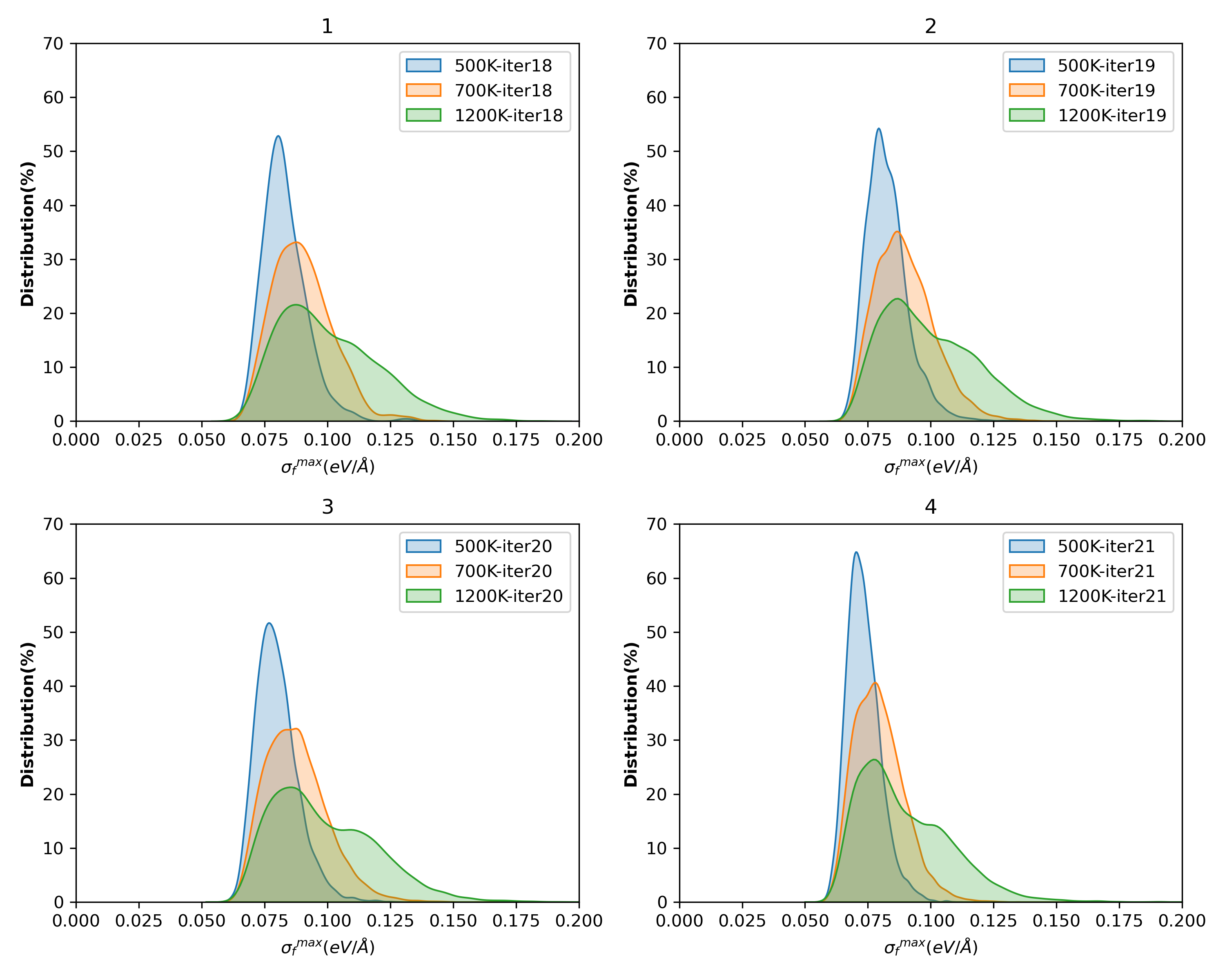}
        \caption{}
        \label{fig:sur_max_devi}       
    \end{subfigure}
    \caption{(a) For bulk Li$_6$PS$_5$Cl, distribution of maximum deviation of force ($\sigma_f^{max}$ ) from iteration 0–11. The distributions of deviation values at four temperatures are plotted. (b) For model trained including surface data of Li$_6$PS$_5$Cl, distribution of maximum deviation of force ($\sigma_f^{max}$) from iteration 18–21. The distributions of deviation values at three temperatures are plotted.}
    \label{devi}
\end{figure}

\begin{figure}[ht]
\centering
    \begin{subfigure}{0.3\linewidth}
        \centering
        \includegraphics[width=1.0\linewidth]{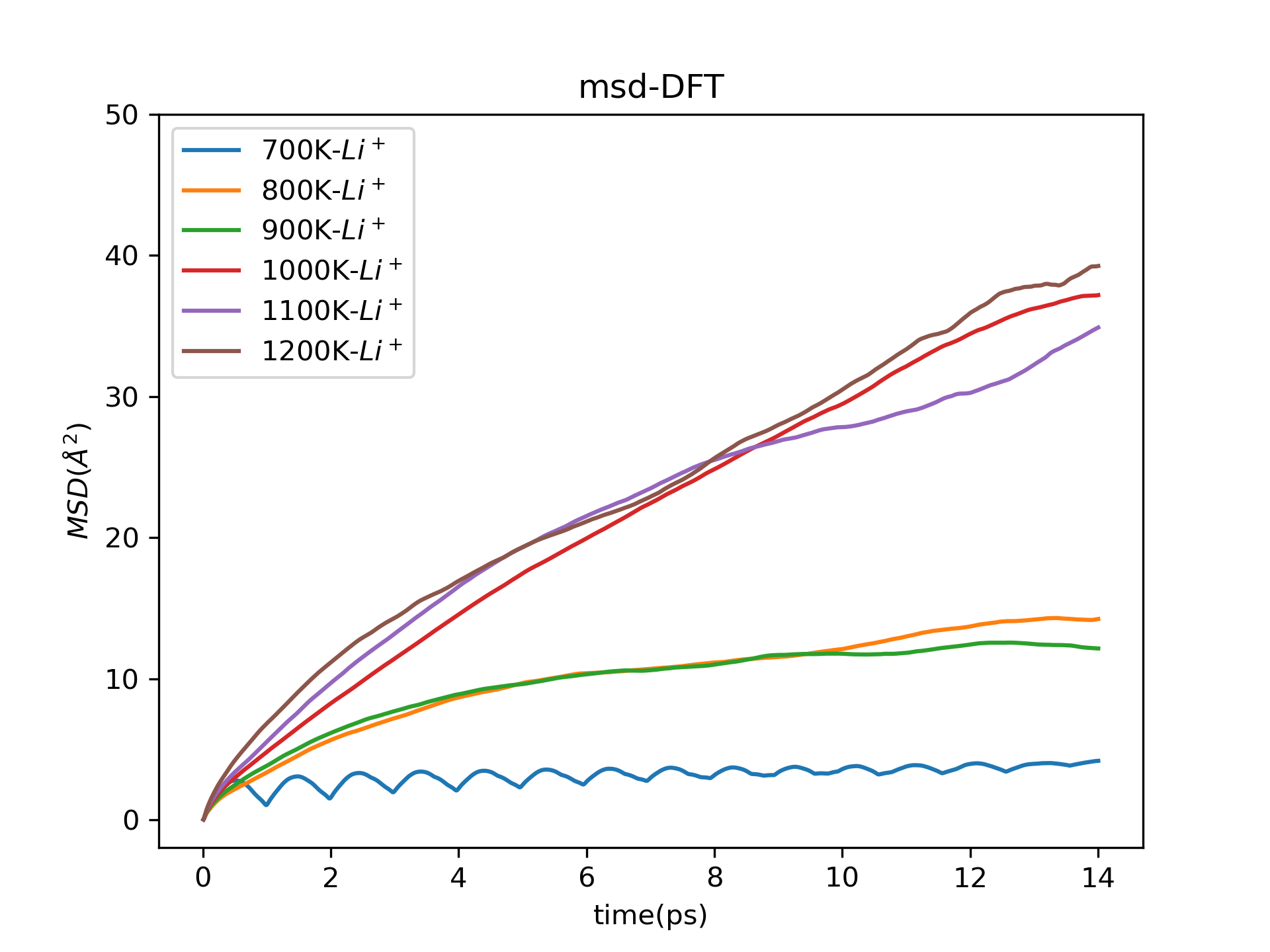}
        \caption{}
        \label{fig:abacus-msd}  
    \end{subfigure} 
    \begin{subfigure}{0.3\linewidth}
        \centering
        \includegraphics[width=1.0\linewidth]{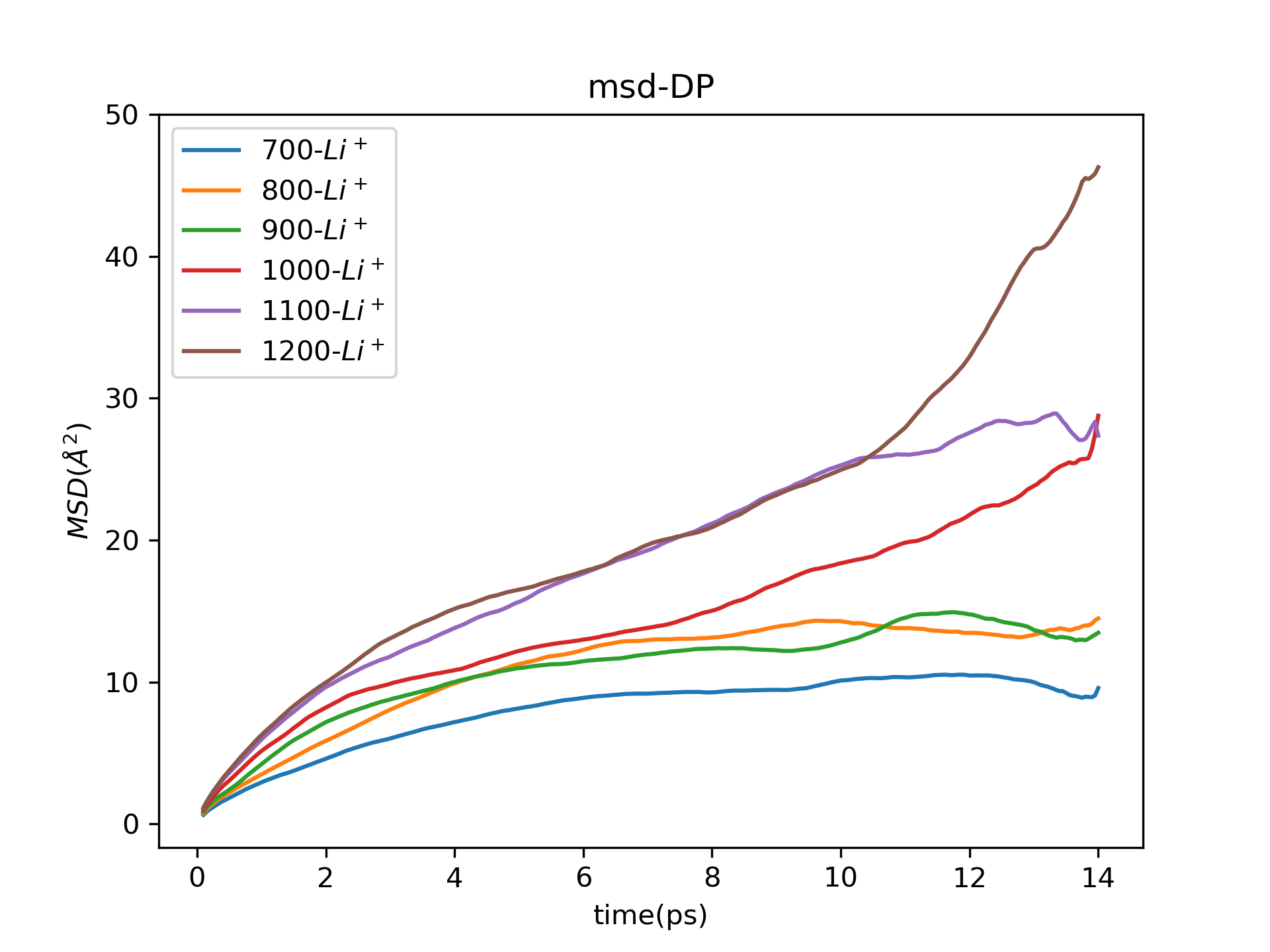}
        \caption{}
        \label{fig:msd-dp}       
    \end{subfigure}
        \begin{subfigure}{0.3\linewidth}
        \centering
        \includegraphics[width=1.0\linewidth]{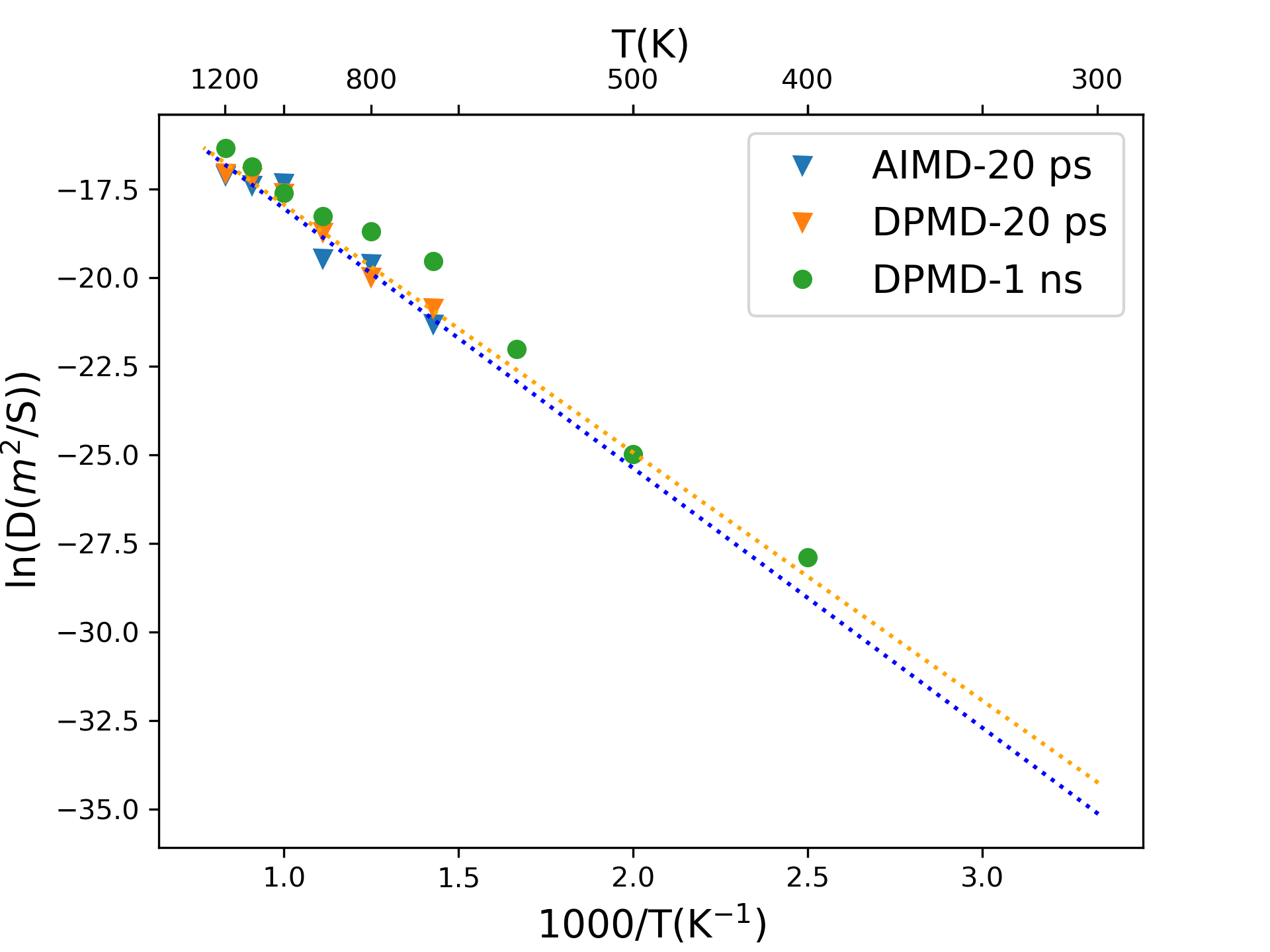}
        \caption{}
        \label{fig:lnd-T}       
    \end{subfigure}
  \caption{ (a) and (b) are the MSD results of AIMD and DPMD simulations for unit cell with time duration of 20 ps. (c)The blue and orange triangles represent the diffusion coefficients evaluated by AIMD and DPMD with the simulation time of 20 ps, and the green dots are obtained from DPMD simulations with the simulation time of 1 ns. }
  \label{fig:MSD}
\end{figure}

The Root Mean Square Error (RMSE) for energy and force  (Fig.~\ref{fig:RMSE}), as well as the equation of state(EOS) curves(Fig.~\ref{fig:EOS}) are used to test the accuracy of DP models. 
Fig.~\ref{fig:max_devi} and Fig.~\ref{fig:sur_max_devi} show the distribution of $\sigma_f^{max}$ at different temperatures in four rounds of training. The first figure of Fig.~\ref{fig:max_devi} contains 200K and 500K data from iter 0-3, 700K data from iter 4-7, and 1200K data from iter 8-11, represents the distribution of $\sigma_f^{max}$ of the first round at these temperature. Four rounds of calculations were conducted for each temperature, making the maximum deviation of the force converge within the criterion. The surface data is included from iter 15-21. The distribution of force deviation for four rounds of training with data from iter 18-21 is given in Fig.~\ref{fig:sur_max_devi}. And the convergence criterion achieves as well. The MSD results from AIMD and DPMD methods simulating within the range from 700K to 1200K are illustrated in Fig.~\ref{fig:MSD}. They have the same simulation time of 20ps, with the first 5ps used for balance and 6-20ps fitting the Arrhenius relationship. The effect of simulation time is shown in Fig.~\ref{fig:lnd-T} by comparing the diffusion coefficients extracted from DPMD simulations with time period of 20 ps and 1 ns. 

\section{Effect of finite-size}
 To analyze the size effect of modeling on the diffusion coefficients, we conducted 5 ns DPMD simulations across various system sizes—52 atoms, 208 atoms, 416 atoms, and 1404 atoms. Observations reveal a noticeable convergence in the diffusion coefficient at a system size of $2 \times 2 \times 2$ within a specific time frame. Balancing accuracy and efficiency considerations, we’ve selected the $2 \times 2 \times 2$ scale for all subsequent simulations.
\begin{figure}[htbp]
    \centering
    \begin{subfigure}{0.4\linewidth}
        \centering
        \includegraphics[width=1.0\linewidth]{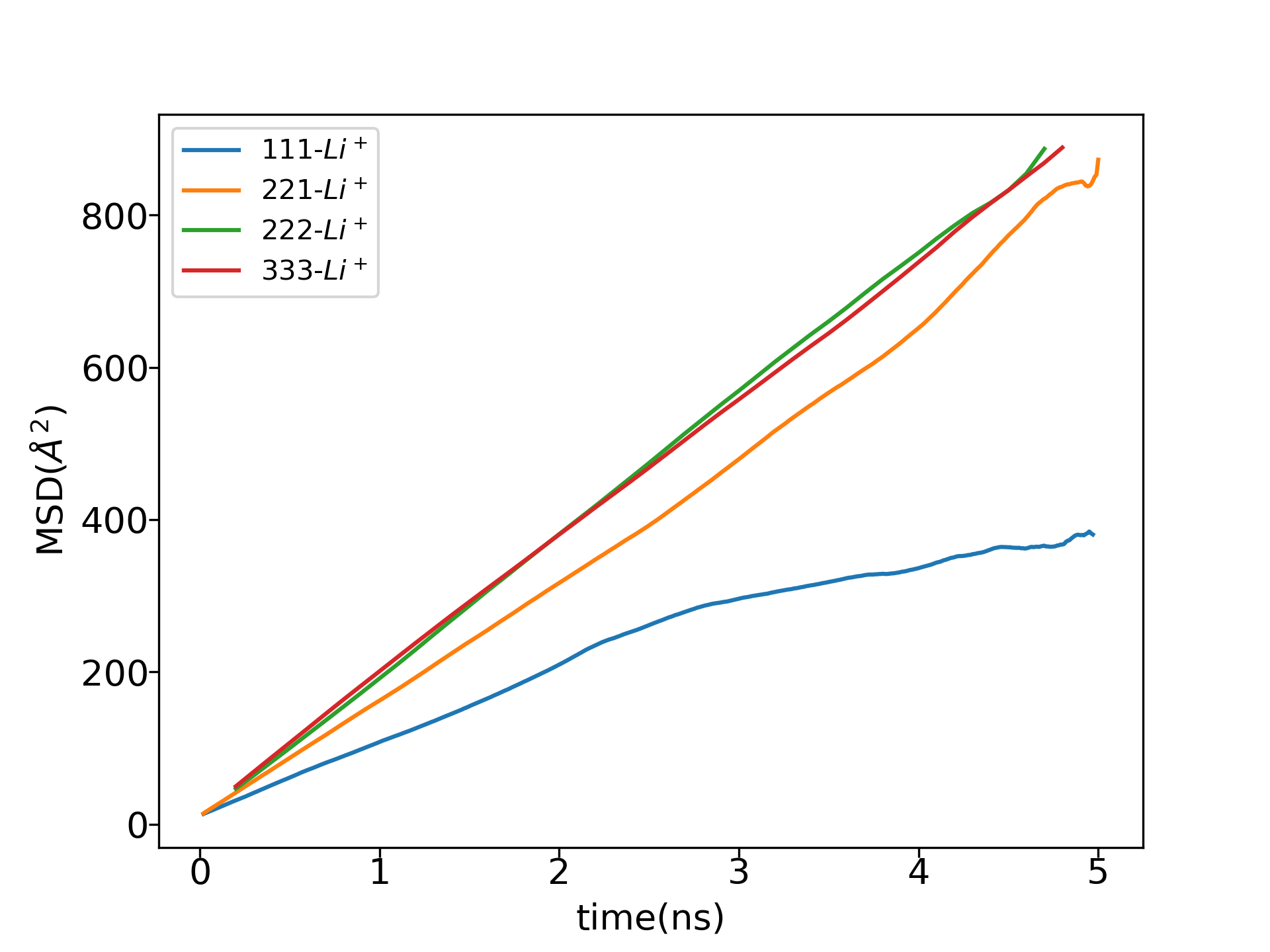}
        \caption{700K}
        \label{fig:700K}  
    \end{subfigure} 
    \begin{subfigure}{0.4\linewidth}
        \centering
        \includegraphics[width=1.0\linewidth]{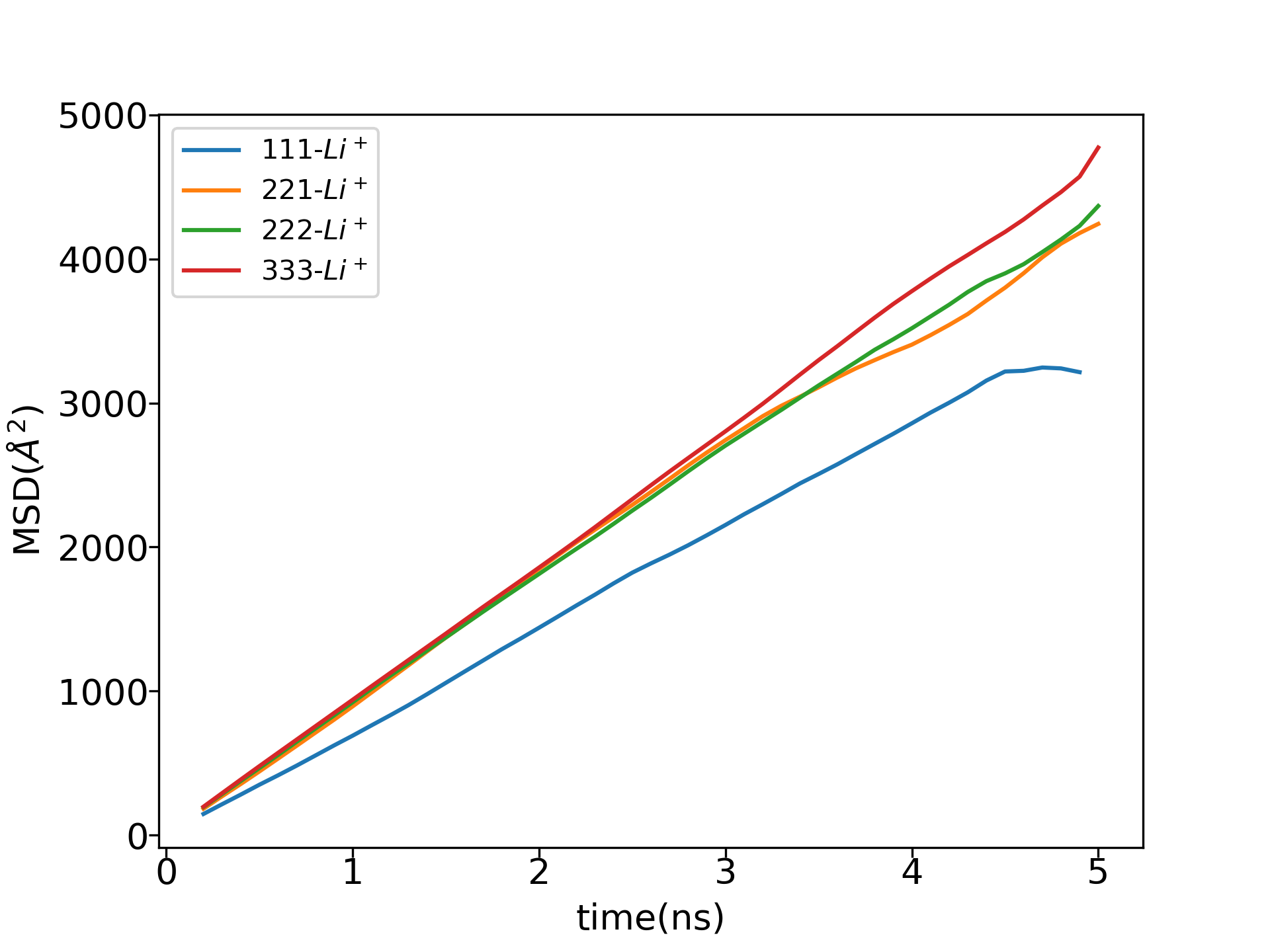}
        \caption{900K}
        \label{fig:900K}       
    \end{subfigure}
    \caption{(a) and (b) are MSD curves of DPMD simulations at 700K and 900K with different cell sizes, including unit cell(52 atoms),  $2\times 2\times 1$ supercell(208 atoms), $2\times 2\times 2$ supercell(416 atoms) and  $3\times 3\times 3$ supercell(1404 atoms) . }
    \label{finite-size}
\end{figure}

\section{Effect of thermal expansion}
\begin{figure}[ht]
    \centering
    \includegraphics[width=0.8\textwidth]{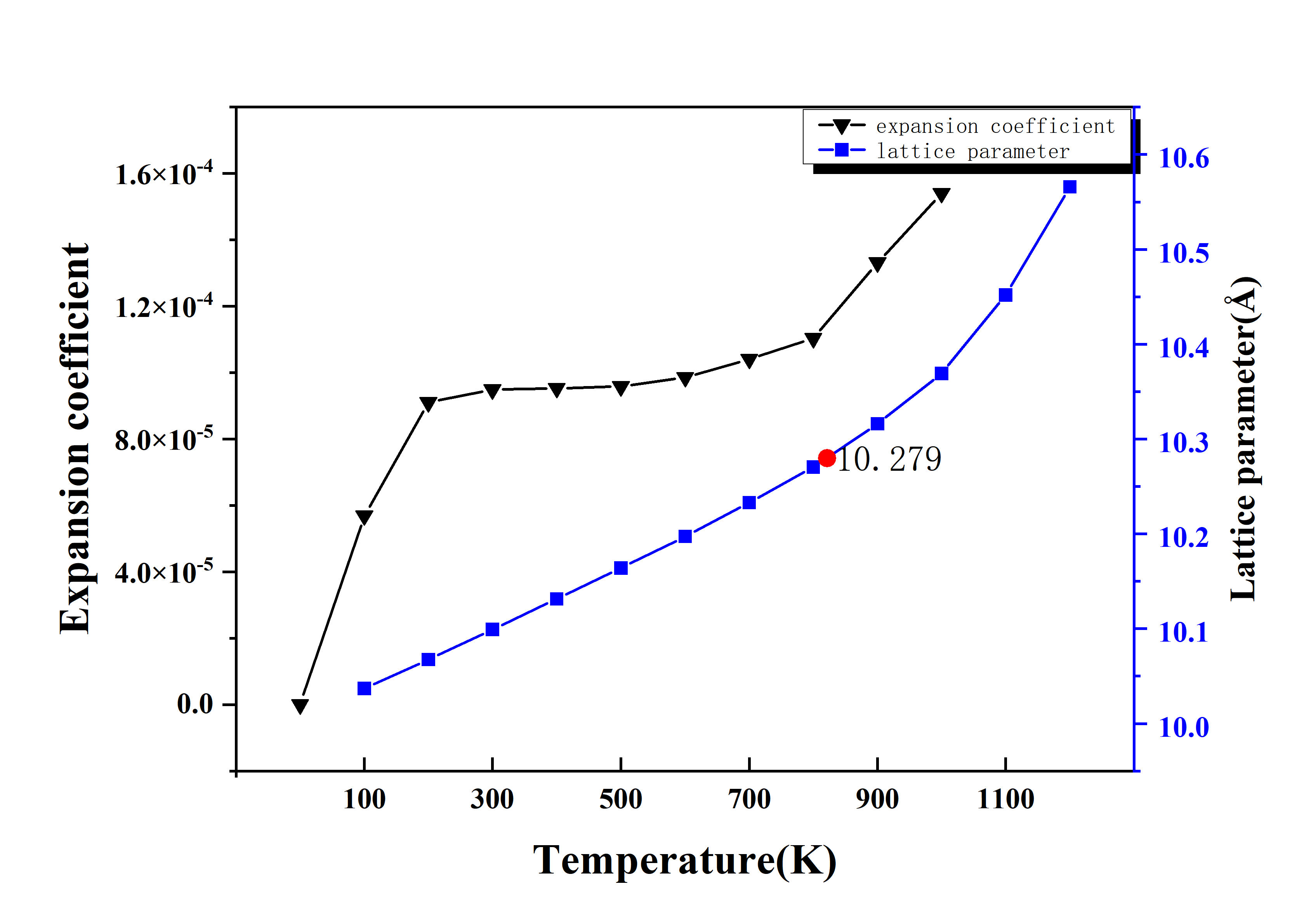}
     \caption{The lattice parameters and the thermal expansion coefficients evaluated by NpT ensemble of DPMD simulations.}
     \label{fig:change-lattice}
\end{figure}

The DP potential also presents reasonable predictive capabilities in lattice scaling, indicating its good extrapolative properties. Then we discussed the influence of lattice expansion or contraction on calculated diffusion coefficients at various temperatures. Fig.~\ref{fig:change-lattice} illustrates the noticeable increase in lattice constants within LPSC as temperature rises. To derive the lattice constants across different temperatures, we stabilized the simulation cell using the NpT ensemble for 2 ns at each temperature. 
The cell volume is averaged over the configurations of the last 0.5 ns to obtain the lattice parameters and thermal expansion coefficients for each simulated temperature. 

\begin{figure}[ht]
\begin{center}
\includegraphics[width=0.8\textwidth]{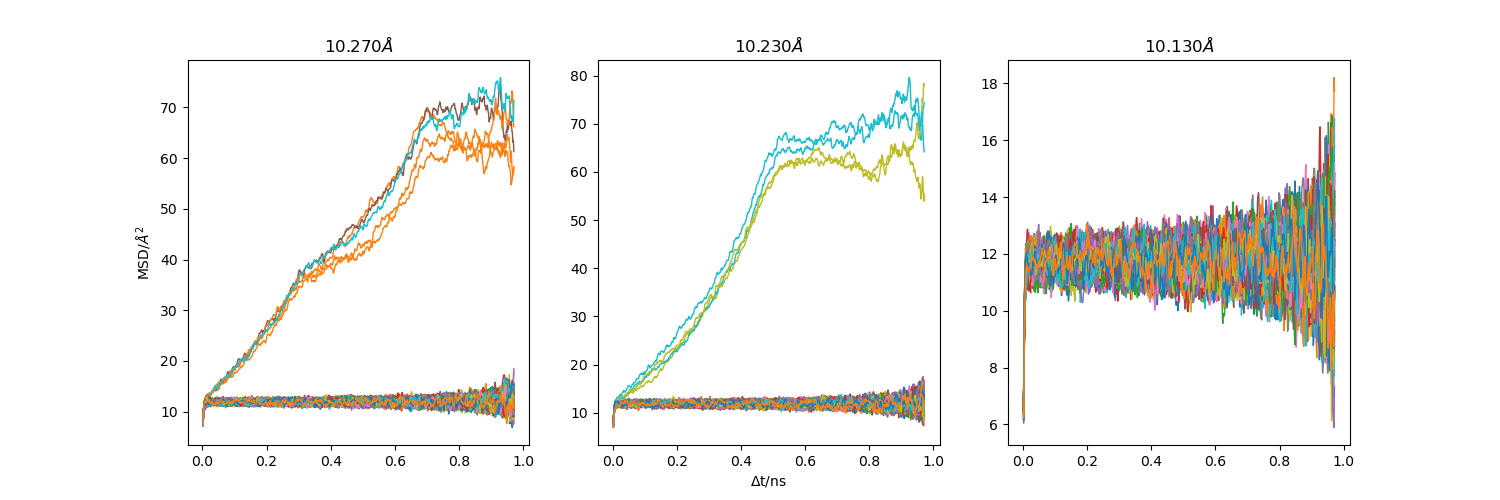}
\caption{The MSD curve for each Li$^+$ ion by DPMD simulations at 400K for 1 ns with three different lattice parameters, 10.279 \AA, 10.230 \AA, and 10.130 \AA. The MSD statistical analysis is averaged to time intervals. }
\label{fig:400-size} 
\end{center}
\end{figure}

\begin{figure}[ht]
\begin{center}
\includegraphics[width=0.8\textwidth]{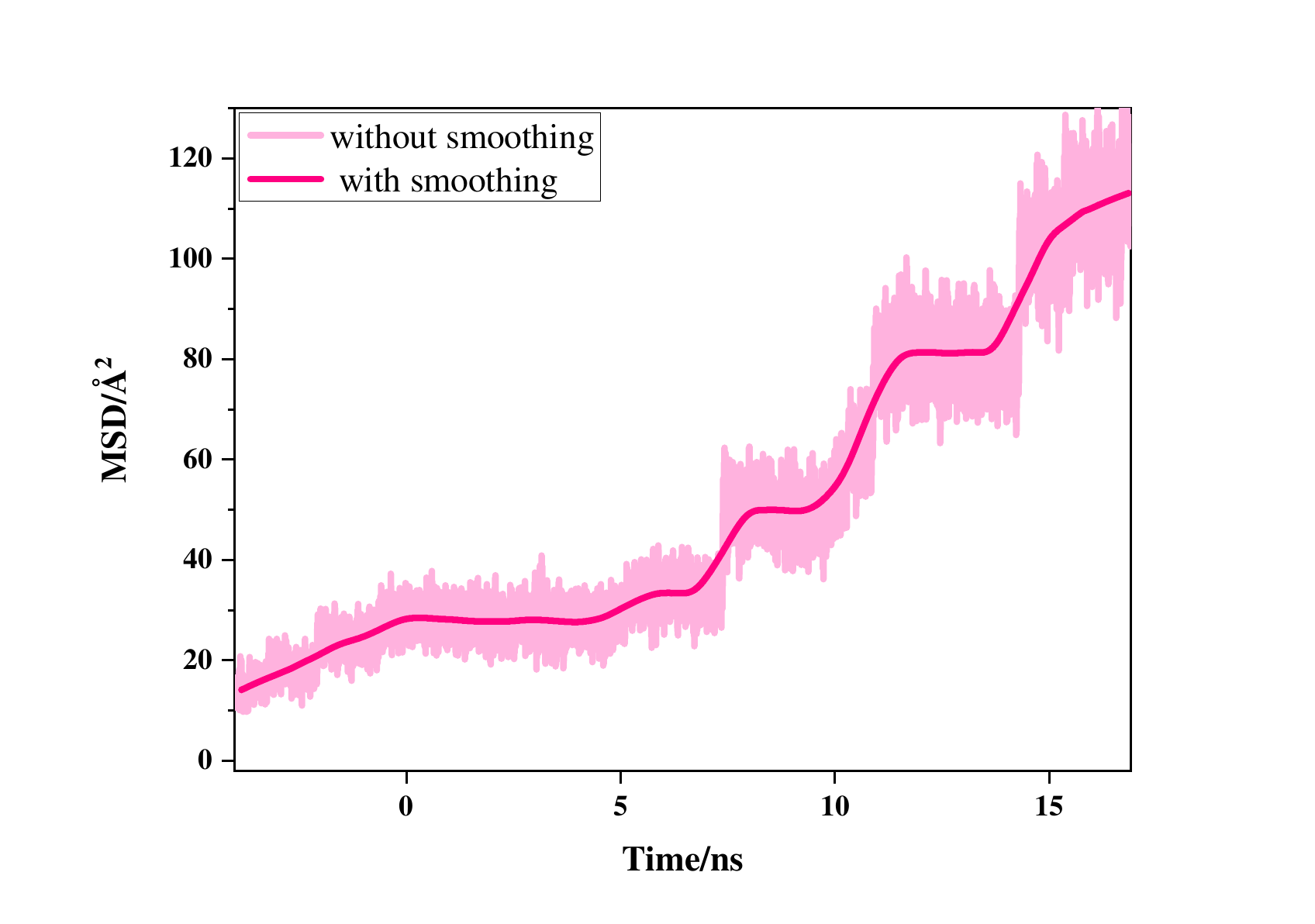}
\caption{The MSD curves for one Li$^+$ ion without averaging to time intervals for DPMD simulation at 500~K with lattice parameter of 10.279 \AA. The number of intercage hopping events is counted by the plateaus appearing along the time evolution.}
\label{fig:NO-per_time} 
\end{center}
\end{figure}

When we adjust the simulated cell lattice at 400K, we will find that as the lattice parameters decrease, the number of Li$^+$ that undergoes migration decreases as well, as shown in Fig.~\ref{fig:400-size},in which the MSD of each Li$^+$ ion in the model is given.
As depicted in Fig.~\ref{fig:NO-per_time}, the MSD analysis at 500~K reveals a pattern of ion movement modes oscillating between intracage and intercage motions. The plateau observed in the figure signifies the rotational motion of Li$^+$ ions around the center of the cage. It provides a straightforward indication of the intercage transition once a new plateau appears along with time evolution, thus the the number of intercage hopping events can be counted at each specific temperature. 

\section{THE EXPLORATION STRATEGY}
The Table~\ref{tab:my_label} shows the details of the exploration strategy, corresponding to Fig.~\ref{fig:max_devi}.
\begin{table}[ht]
    \centering
    \caption{Details of the exploration strategy. For each iteration, we list the initial structures whether have stretch and compression from the equilibrium lattice, the length of trajectories, the simulation temperatures, and the statistical ensembles. For those NpT simulations, 1,50,100,1000 and 2000 Bar are set as the pressures.}
    \begin{tabular}{p{1cm}p{3cm}p{2cm}p{3cm}p{1cm}}
    \toprule
         Iter & STRU & length(ps)& T(K)&Ensemble\\
         \midrule
         0& 0.9-1.1 unit& 200 & 200, 500& NVT \\
         1& 0.9-1.1 unit& 500 & 200, 500& NVT \\
         2& 0.9-1.1 unit& 800 & 200, 500& NVT \\
         3& 0.9-1.1 unit& 800 & 200, 500& NVT \\
         4& 0.9-1.1 unit& 200 & 700, 1000& NVT \\
         5& 0.9-1.1 unit& 500 & 700, 1000& NVT \\
         6& 0.9-1.1 unit& 800 & 700, 1000& NVT \\
         7& 0.9-1.1 unit& 800 & 700, 1000& NVT \\
         8& 0.9-1.1 unit& 500 & 1200& NVT \\
         9& 0.9-1.1 unit& 800 & 1200& NVT \\
         10& 0.9-1.1 supercell& 500 & 500,700,1200& NVT \\
         11& 0.9-1.1 supercell&800 & 500,700,1200& NVT \\
         12& 0.9-1.1 unit& 800 & 700, 1200& NpT \\
         13& 0.9-1.1 unit& 800 & 700, 1200& NpT \\
         14& 0.9-1.1 unit& 800 & 700, 1200& NpT \\
         15& Surface & 20 & 100, 300& NVT \\
         16& Surface & 50 & 100, 300& NVT \\
         17& Surface & 200 & 100, 300& NVT \\
         18& Surface & 100 & 500,700,1200& NVT \\
         19& Surface & 200 & 500,700,1200& NVT \\
         20& Surface & 500 & 500,700,1200& NVT \\
         21& Surface & 800 & 500,700,1200& NVT \\
         \bottomrule
    \end{tabular}
    \label{tab:my_label}
\end{table}
\FloatBarrier

%% file: title.tex
\title{Mechanistic Insights into Temperature Effects for Ionic Conductivity in Li$_6$PS$_5$Cl}

\author{Zicun Li}
\affiliation{College of Physics
Nanjing University of Aeronautics and Astronautics (NUAA)
Nanjing 211106, China}
\affiliation{Beijing National Laboratory for Condensed Matter Physics, Institute of Physics, Chinese Academy of Sciences, Beijing 100190, China}
\author{Jianxing Huang}
\affiliation{State Key Laboratory of Physical Chemistry of Solid Surfaces, iChEM, College of Chemistry and Chemical Engineering,
Xiamen University, Xiamen 361005, China}
\author{Xinguo Ren}
\email{renxg@iphy.ac.cn}
\affiliation{Beijing National Laboratory for Condensed Matter Physics, Institute of Physics, Chinese Academy of Sciences, Beijing 100190, China}
\author{Jinbin Li}
\email{jinbin@nuaa.edu.cn}
\affiliation{College of Physics
Nanjing University of Aeronautics and Astronautics (NUAA)
Nanjing 211106, China}
\author{Ruijuan Xiao}
\email{rjxiao@iphy.ac.cn}
\affiliation{Beijing National Laboratory for Condensed Matter Physics, Institute of Physics, Chinese Academy of Sciences, Beijing 100190, China}
\author{Hong Li}
\affiliation{Beijing National Laboratory for Condensed Matter Physics, Institute of Physics, Chinese Academy of Sciences, Beijing 100190, China}